\documentclass[apsmath, apssymb]{revtex4} 

\usepackage{euscript}
\usepackage{eufrak}
\begin{document} 
\title{A geometric framework for phase synchronization in coupled noisy 
nonlinear systems}
\author{ J.Balakrishnan* \thanks{E-mail: janaki@lorentz.leidenuniv.nl}\\ 
Instituut-Lorentz for Theoretical Physics, 
Universiteit Leiden,\\
Postbus 9506, 2300 RA Leiden, ~The Netherlands.}
\begin{abstract} 
\vspace{2.5cm}

{\bf P.A.C.S. numbers} : ~05.45.-a, ~05.45.Xt, ~05.40.-a, ~02.90.+p, ~89.75.-k  \\

\vspace{2.5cm}

\noindent {\large {\bf Abstract}}\\ 

\noindent A geometric approach is introduced for understanding the phenomenon 
of phase synchronization in coupled nonlinear systems in the presence 
of additive noise. We show that the emergence of cooperative behaviour through 
a change of stability via a Hopf bifurcation entails the spontaneous appearance 
of a gauge structure in the system, arising from the evolution of the slow 
dynamics, but induced by the fast variables. The conditions for the oscillators 
to be synchronised in phase are obtained. The role of weak noise appears 
to be to drive the system towards a more synchronized behaviour. 
Our analysis provides a framework to explain recent experimental observations 
on noise-induced phase synchronization in coupled nonlinear systems. 

\end{abstract} 
\maketitle 
\vspace{3.1cm}
\rule{200pt}{1pt}

{\footnotesize ~*E-mail: janaki@lorentz.leidenuniv.nl, janaki@rri.res.in}\\ 
\newpage 
\section{Introduction} 

\noindent Synchronization phenomena occur abundantly in nature and in 
day to day life. A few well known examples are the observations in 
coupled systems such as pendulum clocks, radio circuits, swarms of 
light-emitting fireflies, groups of neurons and neuronal ensembles 
in sensory systems, chemical systems, Josephson junctions, 
cardiorespiratory interactions, etc.  
Starting from the observation of pendulum clocks by Huygens, 
a vast literature already exists which studies synchronization 
in coupled nonlinear systems --- in systems of coupled maps as well as 
in oscillators and networks [1 \& references therein]. In recent times, 
different kinds of synchronization have been classified --- mutual 
synchronization, lag synchronization, phase synchronization and 
complete synchronization [1-3].\\ 

\noindent Many of these studies aim to understand the properties 
exhibited by the systems once they synchronize or exhibit phase-locking.
A comprehensive understanding still seems to be lacking when one seeks 
to explain why the systems synchronize.
In this paper we introduce a geometric approach in order to address 
this fundamental issue. We aim to understand here the reason for the 
occurrence of synchronized and phase-locked behaviour in coupled 
nonlinear systems which are subject to weak additive noise.\\ 

\noindent We consider a system of $n$ nonlinear oscillators which are coupled. 
We determine the conditions which the dynamically evolving variables of the 
system must satisfy in order that the various oscillators constituting the 
full system synchronize in phase. Our study at the moment does not include 
chaotic systems. We find that the presence of weak noise 
assists in bringing about phase synchronization.\\

\noindent In section 2, we introduce a geometrical approach to discuss 
coupled dynamics. We have adapted the methods which Wilczek \& Shapere 
developed [4-6] to understand self-propulsion of organisms by shape 
deformations in fluids at low Reynolds number, for discussing the deformations 
and changes in the orbit structure in phase space as the system evolves 
in time. 
In section 3 we discuss the dynamics of coupled oscillatory systems in 
the presence of additive noise at the close proximity of a Hopf bifurcation. 
We show the spontaneous emergence of a non-trivial gauge structure for such a 
system arising from the slow degrees of freedom, and induced by the fast 
variables, and associate it with the geometric approach introduced in section 2.   
In section 4, we obtain the condition required to be satisfied by any two 
oscillators to exhibit phase synchronization \& phase locking. 
Section 5 summarizes the main results of the paper and points out directions 
for future studies.\\
Our work is motivated by the need for a theoretical understanding of 
recent experiments on chemical oscillators [7] and numerical simulations [8] 
which show noise-induced phase synchronization in coupled nonlinear systems. 
(See also [9],[29]). \\    
                                                                                      
\section{The geometry underlying the dynamics of coupled systems}

\noindent In a series of beautiful papers [4-6], Shapere \& Wilczek 
established a geometric framework to discuss the motion of deformable objects 
in the absence of applied external force. We adopt these methods to understand 
a fundamental phenomenon in nonlinear dynamics, namely phase synchronization 
in coupled dynamical systems.\\

\noindent In the first section, the main idea underlying the paper is 
developed, which is based on the approach used in [4-6] for a deterministic 
system. Elaborating along these lines, in the following sections we have 
studied a general system of coupled nonlinear oscillators under the 
influence of additive Gaussian white noise and we find the conditions 
under which the coupled units within the full system can exhibit 
phase-locked behaviour and phase synchronization. \\ 

\noindent We consider a system of ~$n$ coupled generalised oscillators 
~$q(x,t)$~ where the state variables $q$ could in general be functions 
not only of time, but could also depend on a set of additional variables 
~$x$, say spatial variables when there is a metric structure associated
with the variables:
\begin{equation}
\dot {q_i} = f_i(q_1, q_2, \dots, q_n, \mu_j) ~~, ~~~~~~~ i = 1, \dots, n ~, 
~~ j = 1, \dots, p  
\end{equation} 
Thus $q_i$ include also extended systems where the individual elements 
could mutually influence each other through a distance-dependent 
interaction between the elements. For instance in the case of coupled 
chemical oscillators [8], $q_i$ would denote concentrations which 
have spatial dependence. Another example occurs in biological 
information processing where neurons interact in ensembles.

We study the simplest case in which, in the absence of couplings, each 
of the $n$ subsystems admits oscillatory solutions for some parameter 
values $\mu_j$. 
Switching on the mutual coupling between these oscillators results in the 
emergence of a collective behaviour. It is then appealing to view the 
collective behaviour as having arisen as a result of some sort of 
communication between different points in the configuration space. 
Thus one is led to a geometrical description of the system's dynamical 
evolution. 

For simplicity, we restrict ourselves in this paper to the case in 
which the collective dynamics also exhibits limit cycle behaviour 
emerging via one or more Hopf bifurcations. 
The more general situation which includes chaotic orbits for the 
uncoupled dynamics is not considered here. 
We define the configuration space of the full system as the space of 
all possible flow lines and closed paths.  
We consider the situation when there is no external driving force, so 
that the space of all possible contours is the space of oriented contours 
centred at the origin. \\
In the absence of mutual couplings, the space of contours 
consists of oriented closed orbits, each orbit inclined at an angle with 
respect to the other. If we now turn on the mutual couplings between these 
$n$ subsystems gradually, each of the orbits would gradually get deformed, 
going through a sequence of shape changes, and resulting subsequently in 
a net rotation for it. The problem of interest is to link the dynamical 
variables of the system with the net rotation induced by a change of 
shape of the orbits for each of the $n$ oscillators in phase space.\\ 

\noindent The relative orientations of any two contour shapes can be 
compared by fixing coordinate axes for each. Since there exists a 
degeneracy in the possible choice of axes one can make 
at each point in the space of contour shapes, each set of reference frame 
we can choose from being isomorphic to $E_n$,  
a gauge structure is therefore induced in this space which 
facilitates going from one particular choice of axes to another. \\ 

\noindent In [4-6], the problem of self-propulsion at low Reynold's 
number made possible solely through shape deformations was discussed. 
Each choice of reference frame fixed to a shape, which assigned a 
``standard location'' in space for each shape, was associated with the 
motion and location of any arbitrary shape in relation to its standard 
location. 
We follow their methods closely to discuss deformations of the oriented 
contours in the space of contour shapes. \\

\noindent Following [5,6], the sequence of oriented contours $S(t)$ can 
be similarly related to the sequence of the corresponding chosen reference 
standard contour shapes $S_0(t)$ by a rigid displacement ${\cal R}(t)$ :
\begin{equation}
S(t) = {\cal R}(t)S_0(t)
\end{equation}
where in general, an $n$-dimensional motion ${\cal R}$ includes both rotations 
$R$ and a translation $l$ :
\begin{equation}
[R, l] = \left ( \begin{array}{ll}
R ~~~l\\
0 ~~~1\\ 
\end{array} 
\right ) 
\end{equation}
where $R(t)$ is an $n\times n$ rotation matrix and stands for a sequence 
of time-dependent rigid motions. The contour boundaries are parametrized 
by the control parameters $\mu_i$, for each of which the rigid motion 
${\cal R}$ acts on the vector ~${[S_0(\mu), 1]}^T$. 
The physical contours $S(t)$ are invariant under a local change 
\begin{equation}
\tilde S_0 = \Omega[S_0]S_0
\end{equation}
made in the choice of standard contours $S_0$. Then the contour 
shape evolution can be written by combining eqn.(4) with eqn.(2) as : 
\begin{equation}
\tilde S(t) = {\cal R}(t){\Omega}^{-1}(S_0(t))\tilde S_0 
= \tilde {\cal R}(t)\tilde S_0(t)
\end{equation}
or 
\begin{equation}
\tilde {\cal R}(t) = {\cal R}(t){\Omega}^{-1}(S_0(t))
\end{equation}
The temporal change in the sequence of rigid motions can be written as:
\begin{equation}
\frac{d{\cal R}}{dt} = {\cal R} ( {\cal R}^{-1}\frac{d{\cal R}}{dt} ) 
\equiv {\cal R}A  
\end{equation}
where $A$ can be identified with the infinitesimal rotation arising from an 
infinitesimal deformation of $S_0(t)$. 
Eqn.(7) can be integrated to obtain the full motion for finite $t$:
\begin{equation}
{\cal R}(t_2) = {\cal R}(t_1) {\cal P} e^{\int_{t_1}^{t_2} A(t) dt}
\end{equation}
where ${\cal P}$ stands for the path ordered integral, the Wilson line 
integral $W$ : 
\begin{equation}
W_{21} = {\cal P} e^{\int_{t_1}^{t_2} A(t) dt} = 1 + \int_{t_1<t<t_2} A(t)dt 
+  \int_{t_1<t<t'<t_2} \int A(t)A(t')dt dt' + \dots 
\end{equation}
in which the matrices are ordered such that the ones occurring at earlier 
times are on the left.\\
It can be seen from eqns.(5),(6) and (7), that $A$ transforms like a gauge potential:
\begin{equation}
\tilde A = \Omega A {\Omega}^{-1} + \Omega \frac{d{\Omega}^{-1}}{dt}
\end{equation} 
and the Wilson integral transforms as:
\begin{equation}
\tilde W_{21} = \Omega_1 W_{21}{\Omega_2}^{-1} 
\end{equation}
Shapere and Wilczek exploited the invariance of (9) under rescaling of time, 
~$t \rightarrow \tau(t)$, 
the measure scaling as ~$dt \rightarrow \dot\tau dt$, 
~$A \rightarrow A/ \dot\tau$, to rewrite it in a 
time-independent geometric form. \\
This was done [5,6] by defining an abstract vector field $A$ on the tangent 
space to $S_0$. 
The projection $A(t)$ of $A$ at the contour shape  $S_0(t)$ is evaluated 
in the direction 
~$\frac{\delta S_0}{\delta t}$ in which the shape is changing: 
\begin{equation} 
A(t) \equiv A_{\dot S_0}[S_0(t)]
\end{equation}
In terms of these projected vector fields, (8) was rewritten in a 
time-independent form for a given path and independent of the manner in 
which the path is parametrised in the contour shape space as: 
\begin{equation}
R(t_2) = R(t_1) {\cal P} e^{\int_{S_0(t_1)}^{S_0(t_2)} A(S_0) dS_0}
\end{equation}
Each of the components ~$A_i[S_0]$ of $A$ coming from each direction in the 
contour space generates a rigid motion and can be defined in terms of a fixed 
basis of tangent vectors ~$\{w_i\}$ at $S_0$ : 
\begin{equation} 
A_i[S_0] \equiv A_{w_i}[S_0]
\end{equation}
An infinitesimal deformation $s(t)$ of a contour $S_0(t)$ can be represented as:
\begin{equation}
S_0(t) = S_0 + s(t)
\end{equation}
where an expansion of $s(t)$ can be made: 
\begin{equation}
s(t) = \sum_i \alpha_i(t)w_i
\end{equation}
It was shown in [5,6] that for the particular case 
$S_0(t_1) = S_0(t_2)$, i,e., for a closed cycle in which the sequence of 
deformations returns the system to the original contour shape in its 
configuration space, the line integral in eqn.(9) becomes the closed 
Wilson loop which can be simplified to  
\begin{equation}
W = {\cal P} e^{\oint A(t) dt} = 1 + \frac{1}{2} \oint \sum_{i,j} F_{ij}
\alpha_i \dot\alpha_j dt 
\end{equation}
where 
\begin{equation}
F_{ij} = \frac{\partial A_{w_i}}{\partial w_j} - \frac{\partial A_{w_j}}
{\partial w_i} + [A_{w_i},A_{w_j}]  
\end{equation}
The field strength tensor $F_{ij}$ gives the resultant net displacement 
when a sequence of successive deformations is made of $S_0$ around a 
closed path and is thus the curvature associated with the gauge potential. \\ 

In the configuration space of contour shapes, the orbit 
of each of the $n$ subsystems of the full coupled systems of oscillators 
undergoes the shape deformations described above. Because of the mutual 
couplings, the motion in phase space of any one oscillator coordinate is 
inseparably linked with that of any other phase space point which may 
be the coordinate of another oscillator. The deformation and motion in 
the configuration space of the various flow lines and closed paths of 
the entire coupled system can thus be viewed as those on the surface of 
a solid deformable body which is undergoing motion solely due to these 
deformations.\\  

The full system of $n$ oscillators can be represented by an $n$-component 
vector ~$\psi^i , ~~(i=1,\dots n)$ ~in an abstract complex vector space:
\begin{equation} 
\psi = \left ( \begin{array}{ll} 
q_1\\
q_2\\ 
\cdot\\
\cdot\\
q_n
\end{array} 
\right ) 
\end{equation}
A rotation through an angle $\Lambda$ with respect to a chosen axis in 
this internal vector space does not change the state of the full system, 
but just takes one oscillator state $q_i$ to another:
\begin{equation}
q_i \rightarrow \tilde q_i = U(\Lambda)q_i = e^{it^\alpha \Lambda^\alpha}q_i
\end{equation}
where $t^k$ are $k$ number of $n\times n$ matrices and are representations 
of the generators of the transformation group. Each of the $q_i$s represents 
the state of the $i$th oscillator at time $t$. \\ 

There are $n$ independent gauge potentials $A_{w_i}$ corresponding to 
the $n$ independent internal rotations. Any two rotation matrices ~
$U(\Lambda^a)$ and ~$U(\Lambda^b)$ do not commute unless $\Lambda^a$ and 
$\Lambda^b$ point in the same direction. On application of a common input 
to the full system, all the different $n$ oscillators respond to it. 
In this case emergence of a collective behaviour is determined by the 
same gauge potential, although perhaps by different amounts or strengths.\\
In the following section we would link these gauge potentials with 
the dynamically relevant variables of the coupled system. In 
section 4 we will attempt to understand the geometrical basis underlying 
the dynamics of phase synchronization between the oscillators.\\

\section{Coupled nonlinear oscillators subject to fluctuations} 
We now consider the system of $n$ coupled nonlinear oscillators $q_i$ 
subject to additive Gaussian white noise $\xi_i$ in the limit of weak noise: 
\begin{equation}
\dot q_i = f_i(q_1,q_2,\dots,q_n, \mu) + \xi_i ~~, ~~~(i=1, \dots, n)
,~~\mu \in R^p.
\end{equation}
where the noise correlations are defined as:
\begin{equation}
\langle \xi_i(t)\xi_j(t')\rangle = Q\delta_{ij}\delta(t-t')
\end{equation}
The eigenvalues of the linear stability matrix of the coupled deterministic 
system determine the route through which the full system moves towards a 
collective behaviour. A pure imaginary complex conjugate pair of eigenvalues 
at the bifurcation point with the remaining $(n-2)$ eigenvalues having nonzero 
real parts signals a Hopf bifurcation. The orbit structure near the 
nonhyperbolic fixed points $(q_0, \mu_0)$ of eqn.(1) is determined by the 
center manifold theorem. When the system described by (22) undergoes a change 
in stability through a Hopf bifurcation, one obtains a $p$-parameter family 
of vector fields on a 2-dimensional center manifold.\\ 
In this case one observes an emergent common frequency of oscillation for the 
coupled system. 
Such a situation automatically realises frequency synchronization also since 
the Hopf oscillator rotates with a characterisic frequency. If there are 
more than one Hopf bifurcations, clearly it indicates more than one common 
frequency of oscillation and one expects to observe a clustering of the various 
$n$ coupled oscillators around these common characteristic frequencies. \\
We aim to understand in this paper how phase synchronization 
results. \\

The full system of n oscillators changes stability 
as the parameter under consideration takes on different values, and at some 
parameter values undergoes bifurcations. We will study the 
system in the close neighborhood of the bifurcation points where the system 
exhibits critical behaviour. It is in these regimes that the behaviours of 
the individual oscillators gives way to the collective behaviour of the 
entire coupled system of the n oscillators. We employ center manifold 
reduction techniques for the system in the presence of fluctuations and 
perform a separation of variables in terms of fast and slow variables as 
in [10-11], exploiting their dynamical evolution on different time scales. 
A drastic simplification can then be made of the system's dynamics and 
one can write the probability $P(q_i,t)$ for the system to be in a certain 
configuration at time $t$ in the weak noise limit as the product:
\begin{equation}
P(q_i,t) = p(q_f|q_s)P(q_s,t)
\end{equation}
where $q_s$  and $q_f$ are the slow and the fast variables respectively 
of the system. 
The probability $P(q_s,t)$ for the critical variables occurs on a slow time 
scale and is non-Gaussian in nature. \\
The properties of the fast variables depend upon the nonlinearities in 
the system. For instance in the case when the coupled system exhibits a 
cusp bifurcation the fast variable could exhibit non-Gaussian fluctuations 
(depending on the specific nonlinear interaction) as it is coupled to 
the critical variable. It can be shown [10] that the joint probability 
density $p(q_f|q_s)$ is confined to a narrow strip peaked about the center 
manifold. We are interested in the case when the coupled system also exhibits 
self-sustained oscillatory behaviour and makes a transition to limit cycle 
behaviour in the presence of fluctuations. \\
It was shown in [10,11], that for a transition via a Hopf bifurcation, 
$p(q_f|q_s)$ has the time-independent Gaussian form in the $q_f$ variables 
with width which depends upon the slow variables $q_s$: 
\begin{equation}
p(q_f|q_s) = {\big(\frac{\sigma(q_s)}{\pi}\big)}^{1/2}e^{-\sigma(q_s)
{(q_f-{q_f}_0(q_s))}^2}
\end{equation}
where the center manifold is obtained as a power series in $q_s$: 
~$q_f = {q_f}_0(q_s)$. ~The center manifold theorem has been used in [12] 
for providing with a proof for the stability of the synchronised states. 
The enslaved stable modes are the fast variables which follow the dynamics 
of the center (critical) modes. 
We rewrite $f_i(q_1,q_2,\dots,q_n, \mu_i)$ as 
\begin{equation}
f_i(q_1,q_2,\dots,q_n, \mu_i) = -\frac{\delta F(q_1,q_2,\dots,q_n, \mu_i)}
{\delta q_i}
\end{equation} 
The Fokker-Planck equation for the full system is:
\begin{equation}
\frac{dP(q_i,t)}{dt} = \frac{dP(q_s,q_f,t)}{dt} = \frac{\partial}{\partial q_i}
(P(q_i,t)\frac{\partial F}{\partial q_i}) + \frac{\partial^2P(q_i,t)}{\partial q_i^2}
\end{equation} 
Using eqn.(23) we can rewrite this as 
\begin{eqnarray}
\frac{dP(q_s,q_f,t)}{dt} &=& \frac{d}{dt}(p(q_f|q_s)P(q_s,t)) 
= \frac{dp(q_f|q_s)}{dt}P(q_s,t) + p(q_f|q_s)\frac{dP(q_s,t)}{dt} \nonumber\\
&=& -( {H_{FP}}_1(q_f,q_s) + {H_{FP}}_2(q_s,t) )P(q_s,q_f,t)   
\end{eqnarray}
Hence in the close proximity of the bifurcation, the operator $H_{FP}$  can 
be written in a separable form, the part $H_{FP}(q_s,t)$ independent of the 
fast variables. Here 
\begin{equation}
- {H_{FP}}_2(q_s,t) )P(q_s,q_f,t) = \frac{\partial}{\partial q_s}
(P(q_i,t)\frac{\partial F}{\partial q_s}) + \frac{\partial^2P(q_i,t)}{\partial q_s^2}
\end{equation}
We find it convenient to analyse the coupled dynamics in a path integral 
framework. We follow the procedure of Gozzi [13] to recast the system,  
eqns.(21),(27) as a path integral, and define: 
\begin{equation}
\Psi = P(q_f, q_s, t) e^{F(q_i)/2}
\end{equation}
so that (26),(28) can be rewritten as
\begin{equation}
\frac{d\Psi}{dt} = -2{\EuScript H}_{FP}\Psi 
\end{equation}
where 
\begin{equation}
{\EuScript H}_{FP} = -\frac{1}{2}\frac{\partial^2}{\partial q_i^2} + 
\frac{1}{8}{(\frac{\partial F}{\partial q_i})}^2 - 
\frac{1}{4}\frac{\partial^2F}{\partial q_i^2}
\end{equation}
To enable computation of correlation functions within the path integral 
formalism, we introduce $n$ external sources $J_i$ to probe the full coupled 
system so that the partition function $Z[J]$ for the system can be written as 
the time ordered path integral 
\begin{equation}
Z[J] = {\cal N}{\cal T}\prod_i\int Dq_f Dq_s D\xi_i e^{-\frac{1}{Q}
\int J_i(t')q_i(t')dt'} P(q_s,t)p(q_f|q_s) \delta(q_i -{q_i}_{\xi})
e^{-\int \frac{\xi_i^2}{4Q}dt'}
\end{equation} 
where ~${q_i}_{\xi}$ denote the solution of the system of Langevin 
equations (21), ${\cal T}$ denotes time ordering and ${\cal N}$ is the 
normalization constant. From eqns.(21) one can write
\begin{equation}
\delta(q_i -{q_i}_{\xi}) = \delta(\dot q_i - f_i(q_1,q_2,\dots,q_n) - \xi_i)
\left\|\frac{\delta\xi_i}{\delta q_i}\right\| 
\end{equation}
We can rewrite the Jacobian 
~$\left\|\frac{\delta\xi_i}{\delta q_i}\right\|$  of the transformation ~
$\xi_i \rightarrow q_i$  as 
\begin{eqnarray}
\left\|\frac{\delta\xi_i}{\delta q_i}\right\| &=& \det \Big[ \Big(
\delta_{ij}\partial_t - \frac{\partial f_i(q_1,q_2,\dots ,q_n)}
{\partial q_j(t')}\Big) \delta(t-t')\Big] \nonumber\\
&=& \exp \{{\rm tr}\ln \partial_t(\delta_{ij}\delta(t-t') 
- {\partial_{t'}}^{-1}\frac{\partial f_i}{\partial q_j(t')})\}
\end{eqnarray}
The operator ${\partial_{t'}}^{-1}$ satisfies the relation
\begin{equation}
\partial_t G(t-t') = \delta(t-t')
\end{equation}
Then we can rewrite eqn.(34) in terms of the Green function in (35) as:
\begin{equation}
\left\|\frac{\delta\xi_i}{\delta q_j}\right\| =  \exp \left\{{\rm tr} 
\left[ 
\ln \partial_t + \ln \left( \delta(t-t') + G_{ij}(t-t')
\frac{\partial f_i}{\partial q_j(t')} 
\right) \right]
\right\} 
\end{equation}
The system evolves forward in time. Hence 
\begin{equation}
G(t-t') = \theta(t-t') 
\end{equation} 
Using this and expanding the logarithm in the argument of the exponential, 
we can simplify eqn.(36) to
\begin{equation}
\left\|\frac{\delta\xi_i}{\delta q_j}\right\| = e^{{\rm tr}\ln \partial_t} 
e^{\int_0^t dt' \theta(0) \frac{\partial f_i}{\partial q_j(t')}}
\end{equation}
Substituting this back into eqn.(32) and using the mid-point prescription 
~$\theta(0)=1/2$  of Stratonovich, we have 
\begin{equation}
Z[J] = {\cal N}{\cal T}\prod_i\int Dq_f Dq_s D\xi_i 
e^{-\frac{1}{Q}\int_0^t J_i(t')q_i(t')dt'}e^{\frac{1}{2}\int_0^t 
\frac{\partial f_i}{\partial q_j(t')}}  
e^{-\frac{1}{4Q}\int_0^t dt' {\big(\dot q_i - 
f_i(q_1,q_2,\dots,q_n)\big)}^2}
\end{equation} 
Eqn.(39) can be reduced to 
\begin{eqnarray}
Z[J] &=& {\cal N}{\cal T}\prod_i\int Dq_f Dq_s 
e^{-\frac{1}{Q}\int_0^t J_i(t')q_i(t')dt'}e^{-\int_0^t dt' 
[\frac{1}{2}\frac{\partial^2F}{\partial q_i\partial q_j} + \frac{1}{4Q}
{\dot q_i}^2 +  \frac{1}{4Q}{(\frac{\partial F}{\partial q_i})}^2]}  
e^{-\frac{1}{2Q}(F(t)-F(0))} \nonumber\\
&=& {\cal N}{\cal T}\prod_i\int Dq_f Dq_s 
e^{-\int_0^t dt' [ {\cal L}^{FP} + \frac{1}{Q} J_i(t')q_i(t')}
e^{-\frac{1}{2Q}(F(t)-F(0))} 
\end{eqnarray}
where we have defined a Fokker-Planck lagrangian 
\begin{equation}
{\cal L}^{FP}(q_i,\dot q_i,t) = \frac{1}{4Q}{\dot q_i}^2 
+ \frac{1}{4Q}{(\frac{\delta F}{\delta q_i})}^2 + \frac{1}{2}\frac{\delta^2F}
{\delta q_i\delta q_j} =  \frac{1}{4Q}{\dot q_i}^2 
+ f_i^2 - \frac{1}{2}\frac{\delta f_j}{\delta q_i\delta q_i}
\end{equation}
which is related to the Fokker-Planck hamiltonian ${\EuScript H}_{FP}$ defined in 
eqns.(30),(31) through a Legendre transformation: 
\begin{equation}
{\EuScript H}_{FP}(\pi_i,q_i,t) = \pi_i\dot q_i - {\cal L}^{FP}(q_i,\dot q_i,t)
\end{equation}
Here $\pi_i$ are the momenta canonically conjugate to the variables $q_i$:
\begin{equation}
\frac{\delta {\cal L}^{FP}}{\delta \dot q_i} = \pi_i = \frac{1}{Q}\dot q_i
\end{equation}
so that  
\begin{equation}
{\EuScript H}_{FP}(\pi_i,q_i,t) = Q\pi_i^2 + \frac{1}{4Q}{(\frac{\delta F}
{\delta q_i})}^2 
+ \frac{1}{2}\frac{\delta^2F}{\delta q_i\delta q_j}  
\end{equation}
We use these relations in eqn.(40) to write the partition function as
\begin{equation}
Z[J] = {\cal N}{\cal T}\prod_i\int D\pi_iDq_i 
e^{-\int_0^t dt' [{\EuScript H}_{FP}(\pi_i,q_i,t) + \frac{1}{Q} J_i(t')q_i(t')]}
\end{equation}
From eqns.(27),(29),(30) and (31), we see that in the close proximity of the 
bifurcation ${\EuScript H}_{FP}$ can be written in a separable form as
\begin{equation}
{\EuScript H}_{FP}(\pi_i,q_i,t) = {\EuScript H}_{FP}(\pi_f,q_f;\pi_s,q_s,t) = 
{{\EuScript H}_{FP}}_1(\pi_f,q_f;q_s,t) + {{\EuScript H}_{FP}}_2(\pi_s,q_s,t)
\end{equation}
Thus the corresponding ${\cal L}^{FP}$ can also be split up as ~~
${\cal L}^{FP}(q_i,\dot q_i,t)= {\cal L}^{FP}_1(q_f,\dot q_f,t) + 
{\cal L}^{FP}_2(q_s,\dot q_s,t)$. ~Then we can write 
\begin{equation}
Z[J] = {\cal N}{\cal T}\int Dq_fDq_s 
e^{-\int_0^t dt' [{\cal L}^{FP}_1(q_f,\dot q_f) + 
{\cal L}^{FP}_2(q_s,\dot q_s,t) + \frac{1}{Q} J_i(t')q_i(t')]}
\end{equation}
An averaging over the fast degrees of freedom enables the partition function 
to be written in terms of an effective Lagrangian as a function of only the 
slow degrees of freedom. This can be done by first rewriting the fast degrees 
of freedom in action-angle variables ($\theta, I$). 
The emergence of a non Abelian gauge structure can then be seen arising from 
the evolution of the slow dynamics but induced by the fast variables. 
After tracing the origin of the induced gauge potential to the slow dynamics, 
we obtain the conditions necessary to be satisfied in order for the coupled 
elements to be synchronised in phase.\\ 
To begin with, we introduce a generating function ~$S^{(\alpha)}(q_f,I;q_s)$ 
which effects the transformation ~$(q_f, \pi_f) \rightarrow (\theta, I)$ 
to the action angle variables:
\begin{equation}
\frac{\partial S^{(\alpha)}(q_f,I;q_s)}{\partial q_i} = \pi_i ~~~~~; 
~~~~~~~~~\frac{\partial S^{(\alpha)}(q_f,I;q_s)}{\partial I_i} = \theta_i 
\end{equation}
$S^{(\alpha)}(q_f,I;q_s)$ is many-valued and time dependent since the slow 
variables change with time.\\

\noindent The phase space structure associated with adiabatic holonomy 
in classical systems was studied by Gozzi and Thacker [14] through 
Hamiltonian dynamics. We find it useful to employ their methods for our 
study of coupled oscillatory systems in a fluctuating environment. 
Denote ~${\cal H}_1(I,q_s,t) = 
{{\EuScript H}_{FP}}_1(q_f(\theta,I,q_s),\pi_f(\theta,I,q_s),t) $.\\ 
Using the canonical transformation law, ${\cal H}_1$ can be expressed in 
terms of the action-angle variables as:
\begin{equation}
{\bar H}_1(\theta,I,q_s(t)) = {\cal H}_1(I,q_s,t) 
+ \dot {q_s}_l\frac{\partial S^{(\alpha)}(q_f,I;q_s)}{\partial {q_s}_l}
\end{equation} 
Using the methods of [14] and [15-17], we determine the dynamics of the $M$ 
critical slow variables $q_s$ of the system by averaging out the $N$ fast 
variables which influence them: 
\begin{eqnarray}
\langle \langle {\bar H}_1 \rangle \rangle &=& \frac{1}{{(2\pi)}^N}\int d^N\theta 
{\bar H}_1(\theta,I,q_s(t))\nonumber\\
&=& \frac{1}{{(2\pi)}^N}\int d^N\theta ({\cal H}_1(I,q_s) 
+ \dot {q_s}_l\frac{\partial S^{(\alpha)}(q_f,I;q_s)}{\partial {q_s}_l})   
\end{eqnarray} 
where the double angular brackets denote the averaging over all $\theta$: 
~~~$\langle \langle f \rangle \rangle = 
\frac{1}{{(2\pi)}^N}\int d^N\theta f$ . \\
Since $S^{(\alpha)}(q_f,I;q_s)$ is multi-valued, the single-valued function 
\begin{equation}
\zeta(\theta,I,q_s) = S^{(\alpha)}(q_f(\theta,I,q_s), 
\pi_s(\theta,I,q_s),q_s) ~~~~~, ~~~~(0\le \theta\le 2\pi) ~~~. 
\end{equation}
is introduced [18]. We have 
\begin{equation}
\frac{\partial\zeta}{\partial {q_s}_l} = \frac{\partial S^\alpha}
{\partial {q_s}_l}  
+ {\pi_f}_i \frac{\partial {q_f}_i}{\partial {q_s}_l}  ~~~~~.  
\end{equation}
Hence this can be substituted into eqn.(50) to obtain 
\begin{equation}
\langle \langle {\bar H}_1 \rangle \rangle = {\cal H}_1(I,q_s)  
+ \dot {q_s}_l \langle \langle \frac{\partial\zeta}{\partial {q_s}_l} 
- {\pi_f}_i \frac{\partial {q_f}_i}{\partial {q_s}_l} \rangle \rangle
\end{equation}
The total Hamiltonian of the system is given, after performing the 
angle averages by:
\begin{eqnarray}
H_{av}(I,\pi_s,q_s) &=& \langle \langle {\EuScript H}_1(q_f,\pi_f;q_s) 
+ {\EuScript H}_2(q_s,\pi_s)\rangle \rangle = \langle \langle {\bar H}_1(\theta,I;q_s) 
+ {\EuScript H}_2(q_s,\pi_s)\rangle \rangle \nonumber\\ 
&=& {\bar H}(I,\pi_s,q_s) + \dot {q_s}_l \langle \langle 
\frac{\partial\zeta}{\partial {q_s}_l} 
- {\pi_f}_i \frac{\partial {q_f}_i}{\partial {q_s}_l} \rangle \rangle 
\end{eqnarray}
where we have let 
\begin{equation}
{\bar H}(I,\pi_s,q_s) = {\cal H}_1(I,q_s) + {\EuScript H}_2(q_s,\pi_s) 
\end{equation}
The Gibbs partition function in eqn.(45) can be rewritten in terms of the 
fast and slow variables as 
\begin{eqnarray}
Z[J] &=& {\cal N}{\cal T}\int D\pi_fD\pi_sDq_fDq_s 
e^{-\int_0^t dt' [{{\EuScript H}_{FP}}_1(q_f,\pi_f;q_s) 
+ {{\EuScript H}_{FP}}_2(q_s,\pi_s) + 
\frac{1}{Q} (J_s(t')q_s(t') + J_f(t')q_f(t'))]} \nonumber\\ 
&=& {\cal N}{\cal T}\int D\pi_sDq_sDI D\theta  
e^{-\int_0^t dt' [{\cal H}_1(I,q_s) + {\EuScript H}_2(q_s,\pi_s) 
- \dot {q_s}_l{\pi_f}_i \frac{\partial {q_f}_i}{\partial {q_s}_l} 
+ \dot {q_s}_l \frac{\partial\zeta}{\partial {q_s}_l} 
+ \frac{1}{Q}(J_s(t')q_s(t') + J_f(t')\theta(t'))]} \nonumber\\ 
&=& {\cal N}{\cal T}\int D\pi_sDq_sDI D\theta  
e^{-\int_0^t dt' [{\bar H}(I,\pi_s,q_s) 
- \dot {q_s}_l{\pi_f}_i \frac{\partial {q_f}_i}{\partial {q_s}_l} 
+ \frac{1}{Q}(J_s(t')q_s(t') + J_f(t')\theta(t'))]}
\end{eqnarray}
Performing the $\theta$ integration and simplifying the resulting expression, 
we get 
\begin{equation}
Z[J] \approx {\cal N}{\cal T}\int D\pi_sDq_sDI e^{-\int_0^t dt'(H_{av}(I,\pi_s,q_s)
+ \frac{1}{Q}J_sq_s)} 
\end{equation}
We use the Magnus expansion [19,20,21] for expanding the time-ordered integral, 
which gives the final state properties in terms of integrals over the initial 
state ones, to rewrite eqn.(57) as:
\begin{eqnarray}
Z[J] &\approx & {\cal N}\int D\pi_sDq_sDI ~\exp \Big\{ -\int_0^t dt'\Big( H_{av} 
+ \frac{1}{Q}J_sq_s   
+ ~\frac{1}{2}[H_{av}(t'),\int_0^{t'}H_{av}(t_1) dt_1] \nonumber\\
&+& \frac{1}{4}[H_{av}(t'),\int_0^{t'}[H_{av}(t_2), \int_0^{t_2}H_{av}(t_1) 
dt_1] ~dt_2]  + \frac{1}{12}[~[H_{av}(t'),\int_0^{t'}H_{av}(t_2)dt_2], 
\int_0^{t'}H_{av}(t_1) dt_1]
\nonumber\\ 
&+& \dots \Big) \Big\} 
\end{eqnarray}
The variation of parameter(s) $\mu_i$ in time of the system brings about 
change in its stability. The commutator terms in the Magnus expansion hence 
arise on account of this parametric time dependence of the Hamiltonian: 
~~$[H_{av}(t'), H_{av}(t_1)] = [H_{av}(\mu_i(t')), H_{av}(\mu_j(t_1))]$. \\ 
The necessity of the time ordering is also motivated by the work of ref.[22] 
who have shown that the order of arrival of signals at an oscillator in a 
network of pulse-coupled oscillators is crucial in determining changes in 
its phase.\\

Retaining terms only upto the first commutator in the expansion and substituting 
for $H_{av}$ from eqn.(54), we obtain after some simplifications:
\begin{eqnarray}
Z[J] &\approx & {\cal N}\int D\pi_sDq_sDI \exp \Big\{-\int_0^t dt'(\bar H(I,\pi_s,q_s) 
+ \dot q_{s_l}\langle \langle \frac{\partial\zeta}{\partial {q_s}_l}\rangle \rangle  
- \dot q_{s_l}\langle \langle {\pi_f}_i\frac{\partial {q_f}_i}{\partial {q_s}_l} 
\rangle \rangle  + \frac{1}{Q}J_sq_s  \nonumber\\
&-& \frac{1}{2}[\bar H(I,\pi_s,q_s,t'), \int_0^{t'}\bar H(I,\pi_s,q_s,t_1) dt_1] 
+ \frac{1}{2}[\bar H(I,\pi_s,q_s,t'), \int_0^{t'}dt_1 
\dot q_{s_m}\langle \langle {\pi_f}_k\frac{\partial {q_f}_k}{\partial {q_s}_m} 
\rangle \rangle  ] \nonumber\\
&+& \frac{1}{2}[\dot q_{s_m}\langle\langle {\pi_f}_k
\frac{\partial {q_f}_k}{\partial {q_s}_m} \rangle\rangle, 
\int_0^{t'}\bar H(I,\pi_s,q_s,t_1) dt_1] 
- \frac{1}{2}[\dot q_{s_l}\langle\langle {\pi_f}_i
\frac{\partial {q_f}_i}{\partial {q_s}_l} \rangle\rangle, 
\int_0^{t'}dt_1 
\dot q_{s_m}\langle\langle {\pi_f}_k\frac{\partial {q_f}_k}{\partial {q_s}_m} 
\rangle\rangle ] \Big\}
\end{eqnarray}
From here we can define an effective Hamiltonian $H_{eff}$: 
\begin{equation}
H_{eff} = \bar H(I,\pi_s,q_s) 
+ \dot q_{s_l}\langle\langle \frac{\partial\zeta}{\partial {q_s}_l}\rangle\rangle  
- \dot q_{s_l}\langle\langle {\pi_f}_i\frac{\partial {q_f}_i}{\partial {q_s}_l} 
\rangle\rangle 
- \frac{1}{2}[\dot q_{s_l}\langle\langle {\pi_f}_i
\frac{\partial {q_f}_i}{\partial {q_s}_l} \rangle\rangle, 
\int_0^{t'}dt_1 
\dot q_{s_m}\langle\langle {\pi_f}_k\frac{\partial {q_f}_k}{\partial {q_s}_m} 
\rangle\rangle ]
\end{equation}
From a Hamiltonian variational principle, it was shown in [14] from simple 
arguments that the averaged fast motion induces an effective gauge field which 
acts on the slow variables. We follow these arguments closely for the coupled 
system subject to fluctuations near the instability. The variational principle 
gives:
\begin{eqnarray}
\delta S_{eff} &=& \delta \int_0^T dt [{\pi_s}_l\dot {q_s}_l 
- H_{eff}(I,\pi_s,q_s)] = 0 \nonumber\\
&=& \delta \int_0^T dt \Big\{ {\pi_s}_l\dot {q_s}_l - {\bar H}(I,\pi_s,q_s) 
- \dot {q_s}_l\langle \langle \frac{\partial\zeta}{\partial {q_s}_l} - 
{\pi_f}_i \frac{\partial {q_f}_i}{\partial {q_s}_l} 
\rangle \rangle 
- \frac{1}{2}[\dot q_{s_l}\langle\langle {\pi_f}_i
\frac{\partial {q_f}_i}{\partial {q_s}_l} \rangle\rangle, 
\int_0^{t'}dt_1 
\dot q_{s_m}\langle\langle {\pi_f}_k\frac{\partial {q_f}_k}{\partial {q_s}_m} 
\rangle\rangle ]
\Big\} = 0 \nonumber\\
&=& \delta \int_0^T dt \Big\{ [{\pi_s}_l +  
\langle \langle {\pi_f}_i \frac{\partial {q_f}_i}{\partial {q_s}_l} 
\rangle \rangle]\dot {q_s}_l -  {\bar H}(I,\pi_s,q_s) 
+ \frac{1}{2}[\dot q_{s_l}\langle\langle {\pi_f}_i
\frac{\partial {q_f}_i}{\partial {q_s}_l} \rangle\rangle, 
\int_0^{t'}dt_1 
\dot q_{s_m}\langle\langle {\pi_f}_k\frac{\partial {q_f}_k}{\partial {q_s}_m} 
\rangle\rangle ]
\Big\}
\end{eqnarray}
The term having the single-valued function $\zeta$ vanishes since it 
is a total time derivative. \\
Varying $S_{eff}$ with respect to $\pi_s$ and $q_s$, keeping the end-points 
fixed gives:
\begin{eqnarray}
\delta S_{eff} &=& \int_0^T dt \Big\{ \delta {\pi_s}_l\big(\dot {q_s}_l 
- \frac{\partial {\bar H}}{\partial {\pi_s}_l}\big) + \delta {q_s}_l 
\Big[ \big( \frac{\partial}{\partial {q_s}_l}\langle \langle {\pi_f}_i 
\frac{\partial {q_f}_i}{\partial {q_s}_m}\rangle \rangle 
-\frac{\partial}{\partial {q_s}_m}\langle \langle {\pi_f}_i 
\frac{\partial {q_f}_i}{\partial {q_s}_l}\rangle \rangle 
+ \frac{1}{2}[\langle\langle {\pi_f}_i
\frac{\partial {q_f}_i}{\partial {q_s}_m} \rangle\rangle, 
\langle\langle {\pi_f}_k
\frac{\partial {q_f}_k}{\partial {q_s}_l} \rangle\rangle ]
\big)\dot {q_s}_m \nonumber\\
&-& \frac{\partial {\bar H}}{\partial {q_s}_l} - \dot {\pi_s}_l \Big] \Big\}
\end{eqnarray} 
We define as in [14], the quantity in angular brackets as  
\begin{equation}
\langle \langle {\pi_f}_i 
\frac{\partial {q_f}_i}{\partial {q_s}_l}\rangle \rangle = A_l 
\end{equation}
Then ~$\delta S_{eff}=0$ ~~ leads to 
\begin{eqnarray}
\dot {q_s}_l &=& \frac{\partial {\bar H}}{\partial {\pi_s}_l} \nonumber\\ 
\dot {\pi_s}_l &=& - \frac{\partial {\bar H}}{\partial {q_s}_l} 
+ \Big( \frac{\partial A_m}{\partial {q_s}_l} 
- \frac{\partial A_l}{\partial {q_s}_m} + \frac{1}{2}[A_l, A_m] \Big)\dot {q_s}_m 
\end{eqnarray}
As in [14] we can identify $A_l$ with a gauge potential, and a 
curvature tensor ~${\cal F}_{lm}$ can be defined as
\begin{equation}
{\cal F}_{lm} =  \frac{\partial A_m}{\partial {q_s}_l} 
- \frac{\partial A_l}{\partial {q_s}_m} + \frac{1}{2}[A_l, A_m]
\end{equation} 
so that the momenta in (64) can be rewritten as 
\begin{equation}
\dot {\pi_s}_l = -\frac{\partial {\bar H}}{\partial {q_s}_l} 
+ {\cal F}_{lm}\frac{\partial {\bar H}}{\partial {\pi_s}_l} 
\end{equation} 
The commutator terms in the momenta and curvature tensor were absent in ref.[14] 
since the Magnus expansion for the time ordered integral was not used there.\\ 

\noindent It should be noted that these commutator terms in eqns.(61),
(62),(64)-(66) arising 
from terms such as ~$[H(\mu), H(\mu')]$ in eqn.(58) could in general impart a 
non-flat nature to the connection and generate a curvature which is non-trivial 
by effectively generating new parameters which were not present in the original 
Hamiltonian. (See for example [23,24] where in the context of examining the 
connection between classical and quantum anholonomy for some interesting systems 
(in particular the displaced harmonic oscillator [25]), it was shown that for 
time-varying Hamiltonians, the original Hamiltonian must be embedded into a larger 
class for locating the effective parameter space where the Berry phase two-form 
has singularities).\\ 
The appearance of a nontrivial gauge structure in general dynamical systems, 
including classical systems, due to a slow variation of the parameters was 
also explicitly demonstrated in the seminal work in [26]. \\

\noindent Eqn.(64) shows that the curvature tensor ${\cal F}_{lm}$ exerts 
a velocity dependent force on the slow variables.  
In order to write canonical equations of motion, one has to therefore 
introduce modified Poisson bracket relations in the slow-variable space: 
\begin{equation}
\{f(q_s,\pi_s), g(q_s,\pi_s)\} = \{ \frac{\partial f}{\partial {\pi_s}_l}
\frac{\partial g}{\partial {q_s}_l} - \frac{\partial f}{\partial {q_s}_l}
\frac{\partial g}{\partial {\pi_s}_l} \} - {\cal F}_{lm}\frac{\partial f}
{\partial {\pi_s}_l}\frac{\partial g}{\partial {\pi_s}_m} 
\end{equation}

Thus the gauge potential coupled to the slow variables is induced 
by the fast degrees of freedom as is evident from (63), the spontaneous 
appearance of the gauge symmetry being associated with the phase degrees 
of freedom of the center modes. \\ 
The emergence of a gauge structure for the system follows from 
the crucial property of separability of the variables as slow and fast ones 
evolving at different time scales, which results from the slaving principle 
for the stable modes near the bifurcation in a noisy system.  
This leads to the motion and deformation of the closed orbits in the 
configuration space. The rotational symmetries of the sequence of successive 
deformations of each orbit brought about the gauge potential discussed in 
Section 2. The analysis above in the current section shows that this can be 
related to the dynamically evolving variables of the coupled system.\\ 

Having made the correspondence of the gauge potential $A_l$ and the curvature 
tensor ${\cal F}_{lm}$ with the dynamics of the actual coupled system through 
the fast and slow variables, we proceed to examine under what conditions phase 
locked behaviour and full synchronization would occur in a coupled system. \\

\section{Condition for synchronization between the coupled oscillators}
At any instant of time, the phase difference between two oscillators ~$q_1$ 
and $q_2$ located at two different points $x$ and $y$ in the configuration 
space can be found from their inner product: 
\begin{eqnarray}
\cos \theta_y &=& \frac{\big( q_2(y), q_1(y) \big)}{|{\big( q_2(y), 
q_1(y) \big)}|} \nonumber\\
&=& \int d^dy d^dx {\rm Tr} \Big( P(e^{\int_x^y A_{\mu}^\alpha(s)t^\alpha 
ds_{\mu}}) \Big)\frac{\big( q_2(y), q_1(x) \big)}{|{\big( q_2(y), 
q_1(y) \big)}|}
\end{eqnarray}
where $\theta_y$ denotes the angle between the oscillators $q_1(x)$ 
and $q_2(y)$ 
in configuration space, measured at the coordinate $y$. The path-ordered 
Wilson line integral appears in the equation above since $q_1(x)$ must 
be parallelly transported to the coordinate point $y$ in order to compare 
it with $q_2$ located at $y$. ~$P$ denotes the path-ordering. 
Since the $q_i$s are related to each other through a gauge transformation 
in the $n$-dimensional configuration space, this can be rewritten using 
eqn.(20) as 
\begin{eqnarray}
\cos \theta_y &=& \int d^dy d^dx {\rm Tr} \Big( e^{-it^{\beta}\Lambda^\beta} 
P(e^{\int_x^y A_{\mu}^\alpha(s)t^\alpha 
ds_{\mu}}) \Big)\frac{\big( q_1(y), q_1(x) \big)}{|{\big( q_2(y), 
q_1(y) \big)}|}\nonumber\\
&=& \int d^dy d^dx {\rm Tr} \Big( e^{-it^{\beta}\Lambda^\beta} 
P(e^{\int_x^y A_{\mu}^\alpha(s)t^\alpha 
ds_{\mu}}) P(e^{\int_y^x A_{\lambda}^\gamma(p)t^\gamma 
dp_{\lambda}})\Big)\frac{\big( q_1(y), q_1(y) \big)}{|{\big( q_2(y), 
q_1(y) \big)}|}\nonumber\\
&=& \int d^dy d^dx {\rm Tr} \Big( e^{-it^{\beta}\Lambda^\beta}  
P(e^{\int_x^y A_{\mu}^\alpha(s)t^\alpha 
ds_{\mu}}) P(e^{\int_x^y A_{\lambda}^\gamma(p)t^\gamma 
dp_{\lambda}}) \Big)\frac{1}{|{\big( q_2(y), q_1(y)\big)}|}
\end{eqnarray}
since ~$\big( q_1(y), q_1(y) \big) =1$. 
If the angle between $q_1$ and $q_2$ remains constant for all times, then 
the oscillators $q_1$ and $q_2$ would be phase-locked; if the angle 
between them is vanishing for all time, the oscillators would be fully 
synchronized in phase with each other. \\  
We would like to determine the conditions under which the phases of 
any two oscillators in a coupled nonlinear system would be 
locked and fully synchronised. 
Since each $q_i$ is an oscillator, each undergoes periodic dynamics 
in the configuration space.\\
Let $q_1(y)$ after being parallelly transported from coordinate $x$, now 
return to the point $x$ during the course of its temporal evolution.  
We denote the state of this oscillator after completing one orbit and  
returning to $x$ by $q_1'(x)$.  
By the time $q_1$ completes this orbit, $q_2$ would have evolved to 
another point $z$. Hence we would now like to calculate the angle 
between $q_2(z)$ and $q_1'(x)$. \\
We have:
\begin{equation}
q_1'(x) = P(e^{\oint A_{\mu}^\alpha(s)t^\alpha ds_{\mu}})q_1(x)
\end{equation}
$q_2(z)$ can be parallelly transported to $x$ to compare it with $q_1'(x)$: 
\begin{equation}
q_2(x) = P(e^{\int_z^x A_{\mu}^\alpha(s)t^\alpha ds_{\mu}})q_2(z)
\end{equation}
Then the angle between $q_1$ and $q_2$ at $x$ can be calculated:
\begin{eqnarray}
\cos \theta_x &=& \frac{\big( q_2(x), q_1'(x) \big)}{|{\big( q_2(x), 
q_1'(x) \big)}|} \nonumber\\
&=&  \int d^dz d^dx {\rm Tr} \Big( 
\big(P(e^{\int_z^x A_{\mu}^\alpha(s)t^\alpha 
ds_{\mu}})\big)^T P(e^{\oint A_{\nu}^\beta(p)t^\beta 
dp_{\nu}})\Big)\frac{\big( q_2(z), q_1(x) \big)}{|{\big( q_2(x), 
q_1'(x) \big)}|}\nonumber\\
&=&  \int d^dz d^dx {\rm Tr} \Big( 
\big(P(e^{\int_z^x A_{\mu}^\alpha(s)t^\alpha 
ds_{\mu}})e^{it^\beta\Lambda^\beta}\big)^T P(e^{\oint A_{\nu}^\beta(p)t^\beta 
dp_{\nu}})\Big)\frac{\big( q_1(z), q_1(x) \big)}{|{\big( q_2(x), 
q_1'(x) \big)}|} \nonumber\\
&=&  \int d^dz d^dx {\rm Tr} \Big( 
\big(P(e^{\int_z^x A_{\mu}^\alpha(s)t^\alpha 
ds_{\mu}})e^{it^\beta\Lambda^\beta}P(e^{\int_x^z A_{\kappa}^\alpha(l)t^\alpha 
dl_{\kappa}}) \big)^T P(e^{\oint A_{\nu}^\beta(p)t^\beta 
dp_{\nu}})\Big)\frac{\big( q_1(x), q_1(x) \big)}{|{\big( q_2(x), 
q_1'(x) \big)}|}
\end{eqnarray}
The change in the angle between $q_1$ and $q_2$ during a time interval $t$ 
can be found using eqns.(69) and (72) and by simplifying the resulting 
expression, to lowest order in $\Lambda$ to be
\begin{equation}
\cos \theta_y - \cos \theta_x = - \frac{C(p)}{2}\int d^dx d^dy \Big\{  
-i\Lambda^aF_{ij}^a + F_{ij}^a\int_y^x A_{\mu}^a(s)ds_{\mu} + 
(\delta^{\alpha a} + \frac{\epsilon^{\beta\alpha a}}{2}\Lambda^{\beta}) 
F_{ij}^a\int_y^x A_{\mu}^{\alpha}(s)ds_{\mu} + \dots \Big\} 
\end{equation}
In arriving at this expression we have used the relation for the Wilson loop 
integral  
\begin{equation}
P(e^{\oint A_{\mu}^\beta(s)t^a ds_{\mu}})= e^{F_{\mu\nu}}
\end{equation}
in which the $t^a$ (introduced earlier in eqn.(20)) are generators of the Lie algebra 
\begin{equation}
[t^a, t^b] = i\epsilon^{abc}t^c
\end{equation}
and 
\begin{equation}
t^aF_{\mu\nu}^a = \partial_{\mu}t^a A_{\nu}^a - 
\partial_{\nu}t^a A_{\mu}^a - i[t^aA_{\mu}^a, t^bA_{\nu}^b] 
\end{equation}
is the curvature tensor of the complex abstract vector space. Also we have 
used the matrix identity 
\begin{equation}
e^Ae^B = e^{A+B + \frac{1}{2}[A,B] + \dots }
\end{equation} 
and the trace relations:
\begin{eqnarray}
{\rm tr}(t_p^a) = 0 \nonumber\\
{\rm tr}(t_p^at_p^b) = C(p)\delta^{ab}
\end{eqnarray}
where $C(p)$ is a constant for the representation $p$. 
As the system we are considering is subject to fluctuations and is not 
deterministic, the quantity which is actually of interest to us is the noise 
average ~$\langle \cos \theta_y - \cos \theta_x \rangle$ ~of the phase 
difference between $q_1$ and $q_2$:
\begin{equation}
\langle \cos \theta_y - \cos \theta_x \rangle = 
-\frac{C(p)}{2} \langle \int d^dx d^dy \Big\{  
-i\Lambda^aF_{ij}^a + F_{ij}^a\int_y^x A_{\mu}^a(s)ds_{\mu} 
+ (\delta^{\alpha a} + \frac{\epsilon^{\beta\alpha a}}{2}\Lambda^{\beta}) 
F_{ij}^a\int_y^x A_{\mu}^{\alpha}(s)ds_{\mu} + \dots \Big\}\rangle  
\end{equation}
Using the Gauss-Bonnet theorem, we see that the first integral on the right 
hand side of this equation gives a topological invariant, the Euler 
characteristic ~$\chi_E$ of the surface ~$S$ over which the integration 
is performed: ~~$\int_S F_{ij} = \chi_E $. 
~~For the situation in which there 
is perfect phase synchronization between any two oscillators in the system, 
this constant term on the right-hand side of eqn.(79) should vanish and 
the other terms in the equation must also vanish. The two-torus $T^2$ is a 
well known example of a topological space with vanishing Euler characteristic.
The limit cycles of the coupled system are therefore constrained to remain 
on $T^2$ as they synchronize in phase.\\
For the oscillators $q_1$ and $q_2$ to exhibit phase-locked behaviour, we 
observe that we must have, to lowest order in $\Lambda$, 
\begin{equation}
\frac{C(p)}{2} \langle \int d^dx d^dy \Big\{  
F_{ij}^a (2\delta^{\alpha a} + \frac{\epsilon^{\beta\alpha a}}{2}
\Lambda^{\beta})\int_y^x A_{\mu}^{\alpha}(s)ds_{\mu} \Big\}\rangle = 
{\rm constant}
\end{equation}
While the explicit value of the left hand side of (80) would vary from one 
set of coupled systems to another, it is interesting to note that in all cases  
external noise seems to play a role in bringing about the phase synchronization.  
This can be seen as follows. The noise averages of $A_{\mu}$ and $F_{\mu\nu}$ 
are calculated by solving the coupled Langevin equations in (21) using eqns.(63) 
and (65). From eqns.(22),(48) and (63) we see that the lowest order terms of 
~$\langle A_{\mu}\rangle$ and $\langle F_{\mu\nu}\rangle$ such as 
~$\langle ~\langle\langle {\pi_f}_i\rangle\rangle ~\rangle$ ~ would not 
contribute so that at least to this order, the left hand side of eqn.(80) 
is brought very close to zero taking the system towards synchrony, 
whereas in the case when no fluctuations are present these terms would 
be non-zero.\\

The role of noise in bringing about phase synchrony can also be understood 
in the following way. 
Consider the analysis by Ermentrout [27] of two weakly coupled oscillators:
\begin{eqnarray}
\frac{1}{\omega_i}\frac{dZ_i}{dt} = F_i(Z_i) + \kappa G_i(Z_i, Z_j) 
~~, ~~i,j=1,2 ~, ~~ i\ne j.  
\end{eqnarray}
$Z_i \in R^{N_i}$, $F_i$ are continuous and differentiable, $G_i$ are 
continuous, and each uncoupled system ~ $\frac{dZ_i}{dt} = F_i(Z_i)$ admits 
a unique, globally stable periodic solution.  
It was shown in [27] that the coupled state admits a parameter regime in 
which $n:m$ phase-locking occurs between the two oscillators, after $n$ 
cycles of oscillator 1 and $m$ cycles of oscillator 2. 
Using a multiple-scale perturbation technique, introducing slow $\tau$ 
and fast $s$  time variables:
~~$\frac{d}{dt} = \omega_2\frac{d}{ds} + \kappa\frac{d}{d\tau}$ ~, 
this $N_1 + N_2$-dimensional 
system was reduced to a one-dimensional evolution equation on a slow 
time-scale for the phase difference $\Phi$ between the two oscillators. This 
was shown to have the form:
\begin{equation}
\frac{d\Phi}{d\tau} = H(\Phi)
\end{equation}
where ~$H(\Phi) = H(\Phi + 2\pi)$, and the phase shifts ~$\Phi$s vary slowly 
in the direction of the flow of the limit cycle which is formed 
due to the coupling of the oscillators. 
Further it was shown that the phase-locked solution to the coupled system 
corresponds to the fixed points of eqn.(82), $H$ being identified with the 
Poincare map for the flow of the full system.\\
For a nonzero value of the noise strength, the sharp transition to the 
critical point is replaced by a bifurcation region, and hence the time spent 
by the unstable modes near the bifurcation (critical slowing down of the 
deterministic system at the bifurcation) is much longer in the stochastic case.\\
It is known [28] that the  slower the system moves along any part 
of the limit cycle, the larger is its statistical weight in that part 
of the limit cycle. Hence, from eqn.(82) and the result of [27] mentioned above, 
the phase-locked solutions of the coupled oscillators are statistically favoured. 
Moreover, since in the stochastic system these (unstable) slowly varying 
phase differences show an increase in the relaxation time as the instability 
is approached, as compared to the deterministic case, phase-locking 
and synchronous solutions have a larger statistical 
weight in the presence of the weak noise.\\ 

We have considered the case in which the coupled system exhibits limit 
cycle behaviour. The formation of a limit cycle involves symmetry-breaking, 
permitting the existence of both stationary and time-dependent probability 
densities (for finite and infinite system volumes respectively). 
The probability peaks for the time dependent densities rotate along the 
limit cycle, while the time-independent densities are crater shaped. 
A particular phase is associated with every given realization of a limit 
cycle and a choice made corresponds to breakdown of gauge symmetry.   
The emergent gauge structure associated with the phase degrees of freedom 
of the center modes in the vicinity of the bifurcation, enables us to introduce  
the geometrical quantities $A_l$ and $F_{lm}$, and to obtain the condition 
for phase-locking (eqn.(80)) in terms of these quantities.\\

\noindent Our analysis was made possible only because of the presence of weak noise 
which, as we showed in Section 3, plays the crucial part of enabling separability 
of the variables near the bifurcation into slow and fast ones evolving 
at different time scales.\\
Measurable phase differences between oscillators  
in the presence of the (weak) Gaussian white noise are noise-averaged 
quantities, and from eqn.(80), are determined largely by the fluctuations 
in the gauge potential. Since these average out to zero, we conclude that 
the presence of (weak) Gaussian white noise always enhances phase synchrony.\\

Recent experimental observations by Fujii {\em et al} [7] of two chemical 
oscillators separated by some distance in the light-sensitive Belousov-Zhabotinsky 
reaction show self-synchronization of phase and frequency by application of noise. 
They observed spontaneous synchronization for small separation distances in the 
absence of noise and demonstrated the existence of an optimum noise intensity for 
the self synchronization phenomenon. Phase synchronization in coupled non-identical 
FitzHugh-Nagumo neurons subject to independent external noise was also demonstrated 
through numerical simulations in [8]. Noise-induced phase and frequency 
synchronization was also demonstrated recently in stochastic oscillatory 
systems both analytically and with numerical simulations [9]. (see also [29]). 

Our analysis provides a framework to understand these findings and opens avenues 
for deeper studies relating cooperative phenomena in coupled nonlinear 
stochastic systems, with the underlying rich geometrical structure of the 
phase space generated by the complex dynamics, and with associated mathematical 
invariants which govern the system's asymptotic behaviour. \\ 

\section{Conclusion}
We have introduced a geometrical approach aiming to understand phase synchronization 
among coupled nonlinear oscillators subject to additive noise. We have considered 
the specific scenario when the collective dynamics of all the oscillators also 
exhibits limit cycle behaviour arising via one or more Hopf bifurcations, 
consequently implying the occurrence of frequency synchronization.
We demonstrate the emergence of a non Abelian gauge structure arising from the 
evolution of the slow dynamics but induced by the fast degrees of freedom. 
The condition required to be satisfied in order for phase locking and phase 
synchronization to be exhibited is obtained in terms of characteristic invariants 
of the surface generated by the dynamics of the system. We find that weak noise 
helps in bringing about phase synchronization. This provides an explanation of 
recent experimental observations and numerical simulations of noise-induced 
phase synchronization [7,8] (see also [9,29]). Our work also motivates 
further studies of the 
internal structure and geometry of synchronization defects in spiral waves in 
oscillatory media which have been areas of keen interest in recent times [30,31].\\

\section*{Acknowledgements}
I am grateful to V.Srinivasan, S.Chaturvedi, Wim van Saarloos, 
and Cornelis Storm for helpful discussions. I would like 
to acknowledge support from the Max Planck Institute for Mathematics in the 
Sciences, Leipzig where most of this work was done, and warm hospitality of 
the Instituut-Lorentz, Universiteit Leiden where it was brought to completion.\\ 

\vspace{\bigskipamount}

\section*{References} 
\begin{enumerate}
\item A.Pikovsky, M.Rosenblum \& J.Kurths, {\em Synchronization: A universal 
concept in nonlinear sciences}, Cambridge University Press, Cambridge (2001).
\item M.G.Rosenblum, A.S.Pikovsky \& J.Kurths, {\em Phys.Rev.Lett.} 
{\bf 76}, 1804 (1996);\\
G.V.Osipov, A.S.Pikovsky, M.G.Rosenblum \& J.Kurths, {\em Phys.Rev.E} 
{\bf 55}, 2353 (1997);\\ 
S.K.Han, T.G.Yim, D.E.Postnov \& O.V.Sosnovtseva, {\em Phys.Rev.Lett.} 
{\bf 83}, 1771 (1999).
\item H.Fujisaka \& T.Yamada, {\em Prog.Theor.Phys.} {\bf 69}, 32 (1983);\\
A.S.Pikovsky, {\em Z.Phys.B} {\bf 55}, 149 (1984);\\
L.M.Pecora \& T.L.Carroll, {\em Phys.Rev.Lett.} {\bf 64}, 821 (1990).
\item A.Shapere \& F.Wilczek, {\em Phys.Rev.Lett.} {\bf 58}, 2051 (1987). 
\item A.Shapere \& F.Wilczek, {\em J.Fluid.Mech.} {\bf 198}, 557 (1989).
\item A.Shapere \& F.Wilczek, in {\em Geometric Phases in Physics}, 
World Scientific Publishing Co, Singapore, (1989).
\item K.Fujii, D.Hayashi, O.Inomoto \& S.Kai, {\em Forma} {\bf 15}, 219 (2000). 
\item B.Hu \& C.Zhou, {\em Phys.Rev.E} {\bf 61}, R1001 (2000). 
\item J-n Teramae \& D.Tanaka, {\em Phys.Rev.Lett.} {\bf 93}, 204103 (2004);\\
D.S.Goldobin \& A.Pikovsky, {\em Phys.Rev.E} {\bf 71}, 045201(R) (2005).
\item E.Knobloch \& K.A.Weisenfeld, {\em J.Stat.Phys} {\bf 33}, 611 (1983).
\item C.Van den Broeck, M.Malek Mansour \& F.Baras, {\em J.Stat.Phys.} {\bf 28}, 
557 (1982);\\
F.Baras,M.Malek Mansour \& C.Van den Broeck, {\em J.Stat.Phys.} {\bf 28}, 577 (1982).
\item H.Haken, {\em Advanced synergetics : instability hierarchies of 
self-organizing systems and devices}, Springer-Verlag, (1983).
\item E.Gozzi, {\em Phys.Rev.D} {\bf 28}, 1922 (1983).
\item E.Gozzi \& W.D.Thacker, {\em Phys.Rev.D} {\bf 35}, 2398 (1987).
\item V.I.Arnold, {\em Mathematical methods of classical mechanics}, 
(Second Edition), Springer-Verlag, (1989). 
\item M.V.Berry, {\em Proc.R.Soc.Lond.A} {\bf 392}, 45 (1984).
\item M.V.Berry, {\em J.Phys.A} {\bf 18}, 15 (1985);\\
J.H.Hannay, {\em J.Phys.A} {\bf 18}, 221 (1985);\\
M.V.Berry \& J.H.Hannay, {\em J.Phys.A}, {\bf 21}, L325 (1988).
\item C.Lanczos, {\em The variational principles of mechanics}, Dover 
Publications, New York, (1986). 
\item W.Magnus, {\em Commun.Pure Appl.Math.} {\bf 7}, 649 (1954).
\item R.M.Wilcox, {\em J.Math.Phys.} {\bf 8}, 962 (1967).
\item R.A.Marcus, {\em J.Chem.Phys.} {\bf 52}, 4803 (1970).
\item M.Timme, F.Wolf \& T.Geisel, {\em Phys.Rev.Lett.} {\bf 89}, 258701 (2002).
\item G.Giavarini, E.Gozzi, D.Rohrlich \& W.D.Thacker, {\em Phys.Rev.D} 
{\bf 39}, 3007 (1989). 
\item G.Giavarini \& E.Onofri, {\em J.Math.Phys.} {\bf 30}, 659 (1989).
\item S.Chaturvedi, M.S.Sriram \& V.Srinivasan, {\em J.Phys.A} 
{\bf 20}, L1071 (1987).
\item F.Wilczek \& A.Zee, {\em Phys.Rev.Lett.} {\bf 52}, 2111 (1984).
\item G.B.Ermentrout, {\em J.Math.Biol.} {\bf 12}, 327 (1981).
\item F.Baras, in {\em Lecture notes in Physics -vol.484}, ed. L.Schimansky-Geier \& T.Poeschel, pg.167, Springer-Verlag, (1997);\\
J.W.Turner, in {\em Proceedings of the International onference on Synergetics}, ed. H.Haken, p.255 (Springer, Berlin), (1980). 
\item L.Callenbach {\em et al.}, {\em Phys.Rev.E} {\bf 65}, 051110 (2002).
\item A.Goryachev, H.Chate \& R.Kapral, {\em Phys.Rev.Lett.} {\bf 80}, 873 (1998);\\
{\it ibid.}, ~{\em Int.J.Bif.Chaos}, {\bf 10}, 1537 (2000).
\item J.Davidsen \& R.Kapral, {\em Phys.Rev.E} {\bf 66}, 055202 (2002).
\end{enumerate}
\end{document}